\documentclass[aps,prl,showpacs,twocolumn,lineno,groupedaddress]{revtex4}  
\usepackage{graphicx}  
\usepackage{dcolumn}   
\usepackage{bm}        
\usepackage{amssymb}   

\newcommand{\MET}{\rlap{\kern0.25em/}E_T}

\begin{document}

\title{Search for First-Generation Scalar Leptoquarks in
$\bm{p \bar{p}}$ collisions at $\sqrt{s}$=1.96 TeV\\ }
\newcommand{\aff}[1]{\ignorespaces $^{#1}$}
\author{
D.~Acosta,\aff{16} J.~Adelman,\aff{12} T.~Affolder,\aff{9} T.~Akimoto,\aff{54}
M.G.~Albrow,\aff{15} D.~Ambrose,\aff{15} S.~Amerio,\aff{42}
D.~Amidei,\aff{33} A.~Anastassov,\aff{50} K.~Anikeev,\aff{15}
% ...
}
\affiliation{
\aff{1}  {\small\it Institute of Physics, Academia Sinica, Taipei, Taiwan 11529,
Republic of China}, \\
\aff{2}  {\small\it Argonne National Laboratory, Argonne, Illinois 60439}, \\
\aff{3}  {\small\it Institut de Fisica d'Altes Energies, Universitat Autonoma
de Barcelona, E-08193, Bellaterra (Barcelona), Spain}, \\
\aff{4}  {\small\it Istituto Nazionale di Fisica Nucleare, University of Bologna,
I-40127 Bologna, Italy}, \\
\aff{5}  {\small\it Brandeis University, Waltham, Massachusetts 02254}, \\
\aff{6}  {\small\it University of California, Davis, Davis, California  95616}, \\
\aff{7}  {\small\it University of California, Los Angeles, Los
Angeles, California  90024}, \\
\aff{8}  {\small\it University of California, San Diego, La Jolla, California  92093}, \\
\aff{9}  {\small\it University of California, Santa Barbara, Santa Barbara, California
93106}, \\
\aff{10} {\small\it Instituto de Fisica de Cantabria, CSIC-University of Cantabria,
39005 Santander, Spain}, \\
\aff{11} {\small\it Carnegie Mellon University, Pittsburgh, PA  15213}, \\
\aff{12} {\small\it Enrico Fermi Institute, University of Chicago, Chicago,
Illinois 60637}, \\
\aff{13}  {\small\it Joint Institute for Nuclear Research, RU-141980 Dubna, 
Russia}, \\ 
\aff{14} {\small\it Duke University, Durham, North Carolina  27708}, \\
\aff{15} {\small\it Fermi National Accelerator Laboratory, Batavia, Illinois
60510}, \\
\aff{16} {\small\it University of Florida, Gainesville, Florida  32611}, \\
\aff{17} {\small\it Laboratori Nazionali di Frascati, Istituto Nazionale di Fisica
               Nucleare, I-00044 Frascati, Italy}, \\
\aff{18} {\small\it University of Geneva, CH-1211 Geneva 4, Switzerland}, \\
\aff{19} {\small\it Glasgow University, Glasgow G12 8QQ, United Kingdom}, \\
\aff{20} {\small\it Harvard University, Cambridge, Massachusetts 02138}, \\
\aff{21} {\small\it Division of High Energy Physics, Department of
Physics, University of Helsinki and Helsinki Institute of Physics,
FIN-00014, Helsinki, Finland}, \\
\aff{22} {\small\it Hiroshima University, Higashi-Hiroshima 724, Japan}, \\
\aff{23} {\small\it University of Illinois, Urbana, Illinois 61801}, \\
\aff{24} {\small\it The Johns Hopkins University, Baltimore, Maryland 21218}, \\
\aff{25} {\small\it Institut f\"{u}r Experimentelle Kernphysik,
Universit\"{a}t Karlsruhe, 76128 Karlsruhe, Germany}, \\
\aff{26} {\small\it High Energy Accelerator Research Organization (KEK), Tsukuba,
Ibaraki 305, Japan}, \\
\aff{27} {\small\it Center for High Energy Physics: Kyungpook National
University, Taegu 702-701; Seoul National University, Seoul 151-742; and
SungKyunKwan University, Suwon 440-746; Korea}, \\
\aff{28} {\small\it Ernest Orlando Lawrence Berkeley National Laboratory,
Berkeley, California 94720}, \\
\aff{29} {\small\it University of Liverpool, Liverpool L69 7ZE, United Kingdom}, \\
\aff{30} {\small\it University College London, London WC1E 6BT, United Kingdom}, \\
\aff{31} {\small\it Massachusetts Institute of Technology, Cambridge,
Massachusetts  02139}, \\
\aff{32} {\small\it Institute of Particle Physics: McGill University,
Montr\'{e}al, Canada H3A~2T8; and University of Toronto, Toronto, Canada
M5S~1A7}, \\
\aff{33} {\small\it University of Michigan, Ann Arbor, Michigan 48109}, \\
\aff{34} {\small\it Michigan State University, East Lansing, Michigan  48824}, \\
\aff{35} {\small\it Institution for Theoretical and Experimental Physics, ITEP,
Moscow 117259, Russia}, \\
\aff{36} {\small\it University of New Mexico, Albuquerque, New Mexico 87131}, \\
\aff{37} {\small\it Northwestern University, Evanston, Illinois  60208}, \\
\aff{38} {\small\it The Ohio State University, Columbus, Ohio  43210}, \\
\aff{39} {\small\it Okayama University, Okayama 700-8530, Japan}, \\
\aff{40} {\small\it Osaka City University, Osaka 588, Japan}, \\
\aff{41} {\small\it University of Oxford, Oxford OX1 3RH, United Kingdom}, \\
\aff{42} {\small\it University of Padova, Istituto Nazionale di Fisica
          Nucleare, Sezione di Padova-Trento, I-35131 Padova, Italy}, \\
\aff{43} {\small\it University of Pennsylvania, Philadelphia,
        Pennsylvania 19104}, \\
\aff{44} {\small\it Istituto Nazionale di Fisica Nucleare Pisa, Universities 
of Pisa, Siena and Scuola Normale Superiore, I-56127 Pisa, Italy}, \\
\aff{45} {\small\it University of Pittsburgh, Pittsburgh, Pennsylvania 15260}, \\
\aff{46} {\small\it Purdue University, West Lafayette, Indiana 47907}, \\
\aff{47} {\small\it University of Rochester, Rochester, New York 14627}, \\
\aff{48} {\small\it The Rockefeller University, New York, New York 10021}, \\
\aff{49} {\small\it Istituto Nazionale di Fisica Nucleare, Sezione di Roma 1,
University di Roma ``La Sapienza," I-00185 Roma, Italy}, \\
\aff{50} {\small\it Rutgers University, Piscataway, New Jersey 08855}, \\
\aff{51} {\small\it Texas A\&M University, College Station, Texas 77843}, \\
\aff{52} {\small\it Texas Tech University, Lubbock, Texas 79409}, \\
\aff{53} {\small\it Istituto Nazionale di Fisica Nucleare, University of Trieste/\
Udine, Italy}, \\
\aff{54} {\small\it University of Tsukuba, Tsukuba, Ibaraki 305, Japan}, \\
\aff{55} {\small\it Tufts University, Medford, Massachusetts 02155}, \\
\aff{56} {\small\it Waseda University, Tokyo 169, Japan}, \\
\aff{57} {\small\it Wayne State University, Detroit, Michigan  48201}, \\
\aff{58} {\small\it University of Wisconsin, Madison, Wisconsin 53706}, \\
\aff{59} {\small\it Yale University, New Haven, Connecticut 06520}}
\collaboration{CDF Collaboration}

\date{\today}

\begin{abstract}

We report on a search for pair production of 
first-generation scalar leptoquarks ($LQ$)
in $p \bar{p}$ collisions at $\sqrt{s}$=1.96 TeV
using an integrated luminosity of 203 $pb^{-1}$
collected at the Fermilab Tevatron collider by the CDF experiment.
We observe no evidence for $LQ$ production
in the topologies arising from $LQ \overline{LQ} \rightarrow eqeq$ and
$LQ \overline{LQ} \rightarrow eq \nu q$,
and derive 95$\%$ C.L. upper limits on the $LQ$ production cross section.
The results are combined
with those obtained from a separately reported 
CDF search in the topology arising from $LQ\bar{LQ} \to \nu q \nu q$
and 95\% C.L. lower limits on the LQ mass as a function of $\beta = BR(LQ \rightarrow eq) $ are derived. 
The limits are 236,  205 and 145 GeV/c$^2$  for  $\beta$ = 1, $\beta$ = 0.5 and $\beta$ = 0.1, respectively.

\end{abstract}


\maketitle

\newpage


The remarkable symmetry between quarks and leptons in the Standard Model (SM) suggests that some more 
fundamental theory may exist, which allows for new interactions between them. 
Such interactions are mediated by a new type of particle,
the leptoquark (LQ)\cite{LQ} and  are predicted in many extensions of the SM, e.g. grand unification,
technicolor, and supersymmetry with R-parity violation\cite{models}. 
A LQ carries both lepton and baryon number, 
is a color triplet boson with spin 0 or 1, and has fractional charge. 
Usually it is assumed that LQs couple to fermions of the same generation  to accomodate experimental constraints 
on flavor changing 
neutral currents and helicity suppressed decays.

Previous experimental limits on LQ production are summarized in \cite{dis2002}.
The H1 and ZEUS experiments at the $e^\pm p$ collider HERA  
published\cite{d02}  lower limits on the mass of a first generation LQ that depend on the unknown LQ $l-q$ Yukawa
coupling $\lambda$. At the LEP collider, pair production of LQs can occur in 
$e^+e^-$ collisions via a virtual $\gamma$ or $Z$ boson in the $s-$channel and lower limits have been 
presented in \cite{d03}. 
At the Fermilab Tevatron\cite{d04,d05,prl7200} LQ would be predominantly pair produced 
through $q\bar q$ annihilation and $gg$ fusion. Since the production is mediated via the strong
interaction it is independent of $\lambda$, in contrast to the searches at {\it e-p} machines.
The coupling strength to gluons is determined by color charges of the particles, and is model-independent in the
case of scalar LQs. The production of vector LQ pairs depends on additional assumptions on LQ coupling to 
gluons and its cross section is typically larger than the cross section for scalar LQs production. 
Since the acceptance for vector and scalar LQ detection is
similar, limits on the vector LQ mass will be more stringent.

In this Letter, we focus on a search for first-generation scalar LQ 
pairs produced in $p \bar{p}$ collisions at $\sqrt{s}$=1.96 TeV.
A search for scalar LQ pairs decaying into 
$\nu\nu qq$, resulting in jets and missing transverse energy topology has been 
presented in \cite{prl7200}. Here we
study alternative final state signatures, with LQs decaying into $eejj$ and a final state
consisting of two electrons and  two jets and LQs decaying in $e \nu jj$ and the final state
consisting of an electron, two jets, and missing transverse energy .  
The results for three channels are combined and presented as a function of $\beta$, 
the LQ branching fraction into an electron and a quark.

CDF is a general--purpose detector built to study
the physics of $p\bar{p}$ collisions at the Tevatron accelerator at Fermilab and it is described in
detail in \cite{CDF4}.
The data used in the analysis were collected during the 2002-2003 Tevatron Run II. The integrated luminosity
for this data sample is 203  $\pm$ 12.2 pb$^{-1}$. Events are selected if they pass the high $E_T$ electron
trigger, requiring one electromagnetic trigger tower to be above threshold and
a set of identification cuts on the electromagnetic cluster, track and shower profile.
The efficiency of the trigger combinations used in the $eejj$ and $e\nu jj$ analyses has been measured
using $Z\to ee$ data\cite{zpaper,muge} and it is $\sim~ 100\% $. 
Electrons are reconstructed offline as calorimeter electromagnetic clusters matching a track in the 
central-tracking system (central electrons, $|\eta|< 1.0$\cite{eta}) or as 
calorimeter electromagnetic clusters only in the forward region ($ 1 \le |\eta| \le 3 $).
Electromagnetic clusters are identified by the characteristics of their energy
deposition in the calorimeter: cuts are applied on the fraction of the energy in the electromagnetic
calorimeter and the isolation of the cluster. 
The identification efficiency for a pair of central electrons is $92.4\% \pm 0.4$
and for a pair of central-forward electrons is $ 80\% \pm 0.4$.
The coordinate of the lepton (also assumed to be the event coordinate )
along the beamline must fall within $60~cm$ of the center of the 
detector ( $z_{vertex}$ cut) to ensure a
good energy measurement in the calorimeter. This cut has an efficiency of $95\% \pm 0.1 (stat) \pm 0.5(sys)$, 
and it was determined from studies with minimum bias events. 
The efficiencies of the identification cuts, the trigger selection and the vertex cut,  
measured using $Z\to ee$ data were taken into account when evaluating the 
signal acceptance and background estimate.
Jets are reconstructed using a cone of fixed radius 
$R = \sqrt{(\Delta\eta^2 + \Delta \phi^2)} = 0.7$ and required to have 
$|\eta| <$ 2.0. 
Jets have been calibrated as a function of $\eta$ and $E_T$
and their energy is corrected to the parton level\cite{topPRD}.
Neutrinos produce missing transverse energy, $\MET$, which is measured by balancing the calorimeter 
energy in the transverse plane.

In the analyses we describe here, the signal selection criteria are set according to the kinematic distribution
(e.g. $E_T$ of the electrons and $E_T$ of the jets) of decay products determined from Monte Carlo (MC) studies, optimized to
eliminate background with a minimal loss of signal events\cite{federicaThesis,danThesis}.
In the dielectron and jets topology, we select events 
with two reconstructed isolated electrons with $E_T > $ 25 GeV.
At least one electron is required to be central, 
while the other can be central or forward.
Events are further selected if there are at least two jets with
$E_T>$ 30 and 15 GeV.
The dataset selected above is dominated by QCD production of $Z$ 
bosons in association with jets and $t\bar t$ production where
both the $W$'s from top decay into an electron and neutrino.
To reduce these backgrounds the following cuts are applied: 
i) veto of events whose reconstructed dilepton mass falls in the window
$76 < m_{ee} < 110 $ GeV/c$^2$ to remove the most of the $Z$ + jets contribution, 
ii) $E_T(j_1) + E_T(j_2) > 85$  GeV  and  $E_T(e_1) + E_T(e_2) > 85$  GeV, 
iii) $\sqrt { ( E_T(j_1) + E_T(j_2) )^2 +  (E_T(e_1) + E_T(e_2) )^2 } >$  200 GeV to remove 
the remaining $Z$ + jets and top contributions.
We studied the properties of the physics backgrounds by 
generating the process $Z$ + 2 jets with 
ALPGEN\cite{alpgen} + HERWIG \cite{HERWIG} (to perform parton showering) 
and $t\bar t$  with PYTHIA \cite{pythia},
then passing them through a complete simulation of the CDF II detector based on 
GEANT\cite{geant} and full event reconstruction. 
Other backgrounds from $b\bar b$, $Z \to \tau \bar \tau$, $WW$ are negligible due to 
the electron isolation and large electron and jet transverse energy requirements. 
To normalize the number of simulated events to data we used the theoretical cross sections for 
$t\bar t$ from \cite{mlm} and for $\gamma / Z \to ee$ + 2 jets from \cite{mcfm}.
The expected number of  $Z$ + 2 jets events is $1.9 \pm 0.4$.  
The expected number of $t \bar t$ events is $0.35 \pm 0.06$ events. 
The background arising from multijet events where a jet is mismeasured as an electron
(fake) is calculated using data, for both this analysis and the one that follows.
The method used relies on the assumption that the fake electron 
produced by a jet will be accompanied by other particles
produced by the fragmentation of the jet; thus the isolation fraction of 
the fake electron will generally be larger than the one corresponding to 
a real electron.
The isolation fraction is defined here as:
$
(E_T^{cone} -E_T^{cluster}) / E_T^{cluster}
$ 
where $E_T^{cone}$ is the sum of the electromagnetic and hadronic
transverse energies measured in all towers in a radius $R=\sqrt{(\Delta\phi^2 +\Delta\eta^2)}= 0.4$ around the
electron and $E_T^{cluster}$ is the transverse electromagnetic energy of the electron.
The phase space corresponding to the two electron isolation fractions ($eejj$)  or to one electron 
isolation fraction and the $\MET$ ($e\nu jj$)
is divided in different regions.
We assume that
there is no correlation between the isolation of the two electrons ($eejj$)
and the isolation of the electron and $\MET$ ($e\nu jj$). 
In the region where both electrons have large isolation fraction ($eejj$), 
or where the $\MET$ is small and the isolation fraction of the electron is large ($e\nu jj$)
the $LQ$ contribution is expected to be negligible. We call these background-dominated regions.
With these assumptions from the ratio of the number of events in the 
background-dominated regions we can extrapolate the contribution
in the signal region.
We estimate $0^{+0.7}_{-0}$ fake events in the central-central category
and $4.0\pm  2.0$ in the central-forward category. The final background estimate is 
$6.2 \pm 2.2$ events. 
We checked the prediction of our background sources with data in a control region defined by requiring
two electrons with $E_T > $ 25 GeV, 2 jets with $E_T >$ 30 GeV and $ 66 < m_{ee} < 110$ GeV/c$^2$. 
We observe 107 events in agreement with 113 $\pm$ 15  predicted from SM processes.

The efficiency to detect our signal  was obtained from 
MC simulated LQ (PYTHIA) to account for kinematical and geometrical acceptances and it is
reported in Table I for various LQ mass values.
\begin{table}
\caption{\label{tab:eff}
Efficiencies after all cuts with total error (statistical and systematic) and 95\% C.L. upper limits on the production cross 
section $\times$ branching fraction Br, as a function of M$_{LQ}$, for the two channels.}
\begin{ruledtabular}
\begin{tabular}{crcrc}
$M_{\rm{LQ}}$ (GeV/$c^2$)  & \multicolumn{2}{c}{$eejj$ } & \multicolumn{2}{c}{$e \nu jj$  } \\
\hline
& \multicolumn{1}{l}{$\ \epsilon$}(\%) & $\sigma\times$Br(pb)
& \multicolumn{1}{l}{$\ \epsilon$}(\%) & $\sigma\times$Br(pb) \\
\hline
     100              &    7 $\pm$ 0.5 & 1.11          &  2 $\pm$ 0.26  &  5.71    \\
     140              &    12 $\pm$ 0.5 & 0.25          & 8 $\pm$ 0.7  &  0.69   \\
     160              &    21 $\pm$ 0.8 & 0.14          & 8 $\pm$ 0.7  &  0.65   \\
     200              &    32 $\pm$ 1.2 & 0.09          & 16 $\pm$ 1.3  &  0.37   \\
     220              &    35 $\pm$ 2.0 & 0.08          & 19 $\pm$ 1.5  &  0.24   \\
     240              &    38 $\pm$ 2.0 & 0.07          & 20 $\pm$ 1.6  &  0.23   \\
     260              &    40 $\pm$ 2.0 & 0.07          & 22 $\pm$ 1.7  &  0.22   \\
  \end{tabular}
\end{ruledtabular}
\end{table}
The following systematic uncertainties are
considered when calculating signal acceptance and background predictions:
luminosity (6\%), 
choice of parton distribution functions (2.1\%), 
statistical uncertainty of MC  ($<$ 1\%), 
jet energy scale  ($<$ 1\%),  
statistics of $Z\to e^+e^-$ sample (0.8\%) and
$z_{vertex}$ cut (0.5\%).
After all selection cuts, 4 events are left in the $ee jj$ channel  data.

In the search in the electron and neutrino plus two jets topology,
we select events with one reconstructed isolated electron with $E_T > $  25 GeV. 
The electron is required to be central ($|\eta| \le 1.0 $). We veto events with a second
central or forward electron to be orthogonal to the previous analysis.
We then select events where there is a large missing transverse energy, $\MET  > 60$ 
GeV and at least two jets with E$_T>$ 30 GeV in the range $|\eta| \le 2 $.
This time the selected dataset is dominated by QCD production of $W$
bosons in association with jets and top quark pairs, where either both the $W$'s
from the top pair decay into $l\nu$ and one lepton is not identified, or one of
the $Ws$ decays leptonically and the other hadronically.
A small source of background is represented by $Z$ + 2 jets, where one of the electrons is not identified. 
To reduce these backgrounds the following cuts are applied: 
i)$\Delta\phi (\MET-jet) > 10^o$ to veto events where the 
transverse missing energy is mis-measured due to a jetpointing to a non instrumented region of the calorimeter , 
ii)$E_T(j_1) + E_T(j_2) > 80~ GeV$, iii) transverse mass of electron-neutrino system, 
$M_T(e\nu) > 120~ GeV/c^2$ to reduce the $W$ + 2 jets contribution.
We studied the properties of the $W$ + jets,  $t\bar t$ and $Z$ + 2 jets backgrounds using MC simulated events 
(ALPGEN + HERWIG and PYTHIA).
The background from $W\to \nu_{\tau} \tau$ + 2 jets (ALPGEN+HERWIG)
is negligible after the final mass window cut (see below), as as is the background coming
from misidentified leptons and false $\MET$.
Our final cut consists in selecting events falling in a mass window defined around the LQ mass in the following way.
We calculate the invariant mass of the electron-jet system and the transverse mass of the neutrino-jet system. 
Given the decay of the two LQs, there are two possible 
mass combinations for the electron and the neutrino with the two leading jets. We choose 
the combinations that minimize the difference between the electron-jet mass and the neutrino-jet transverse mass.
We fit the peak of the {\it e-jet} distribution with a Gaussian, to obtain an estimate of the spread of 
the distribution in the signal region ($\sigma_{e}$), as well as the $\nu -jet$ transverse  mass distribution, 
to obtain $\sigma_{\nu}$. In the kinematic plane of $m(e-jet)$ vs $m_T(\nu -jet)$ we define the 
sides of rectangular boxes centered around various nominal LQ mass as $3\times \sigma_{e,\nu}$. 
For each LQ mass, events are accepted if they fall inside the rectangular box.
The overall selection efficiency for various LQ masses is given in Table I.
We checked the simulation prediction of our background sources with data in a control region 
defined by requiring one electron with $E_T > $ 25 GeV, $\MET >$ 35 GeV and 2 jets with $E_T >$ 30 GeV. 
We observe 536 events in agreement with 503~$\pm$~22 predicted from SM processes.

The efficiency to detect our signal was obtained from 
MC simulated LQ data (PYTHIA).
The following systematic uncertainties are considered when calculating signal acceptance and 
background predictions:
luminosity (6\%), 
choice of the parton distribution functions (2.1\%), 
statistics of MC  ($<$ 1.0\%), 
jet energy scale  ($<$ 1\%),
electron identification (0.6\%), 
$z_{vertex}$ cut (0.5\%), 
initial and final state radiation (1.7\%).
The number of events in each mass region, after all selection cuts, 
compared with the background expectations is reported in Table II.
\begin{table}
\caption{\label{tab:enujjfinal}
Final number of events surviving all cuts in the electron, missing energy and jets topology,
 compared with background expectations, as a function of the LQ mass (in GeV/c$^2$). Errors include statistical
and systematic uncertainty.}
\begin{ruledtabular}
\begin{tabular}{crcrcr}
     Mass             & W + 2 jets      & top             &  Z + 2 jets      & Total            & Data \\
     \hline
     120              & 1.5 $\pm$ 0.9   &  3.3 $\pm$ 0.5  &  0.06 $\pm$ 0.01 &  4.9 $\pm$ 1.0   &  6   \\
     140              & 1.5 $\pm$ 0.9   &  3.1 $\pm$ 0.6  &  0.08 $\pm$ 0.02 &  4.7 $\pm$ 1.0   &  4   \\
     160              & 2.5 $\pm$ 1.1   &  2.8 $\pm$ 0.6  &  0.08 $\pm$ 0.02 &  5.4 $\pm$ 1.2   &  4   \\
     180              & 2.5 $\pm$ 1.1   &  2.4 $\pm$ 0.5  &  0.08 $\pm$ 0.02 &  5.0 $\pm$ 1.2   &  4   \\
     200              & 2.5 $\pm$ 1.1   &  2.0 $\pm$ 0.5  &  0.07 $\pm$ 0.02 &  4.6 $\pm$ 1.2   &  4   \\
     220              & 2.0 $\pm$ 1.0   &  1.6 $\pm$ 0.3  &  0.06 $\pm$ 0.02 &  3.7 $\pm$ 1.1   &  2   \\
     240              & 2.0 $\pm$ 1.0   &  1.1 $\pm$ 0.3  &  0.06 $\pm$ 0.02 &  3.1 $\pm$ 1.0   &  2   \\
     260              & 1.5 $\pm$ 1.0   &  0.8 $\pm$ 0.3  &  0.04 $\pm$ 0.02 &  2.4 $\pm$ 0.9   &  2   \\
  \end{tabular}
\end{ruledtabular}
\end{table}

In the analyses described above the number of events passing the selection cuts is consistent
with the expected number of background events. The conclusion of the two searches is that there is no LQ
signal: hence we derive an upper limit on the LQ production cross section at 95\% confidence level.
We use a Bayesian approach\cite{bayes} with a flat prior for the signal cross section and  
Gaussian priors for acceptance and background uncertainties. 
The cross section limits are tabulated in Table I and the mass limits are tabulated in Table III.
To compare our experimental results with the 
theoretical expectation, we use the next-to-leading order (NLO) cross-section for scalar LQ pair
production from \cite{kramer} with CTEQ4 PDF\cite{cteq}.
The theoretical uncertainties correspond to the variations from $M_{LQ}/2$ to $2M_{LQ}$ of the 
renormalization scale $\mu$ used in the calculation. To set a limit
on the LQ mass we compare our 95\% CL upper experimental limit to the theoretical cross section for $\mu = 2M_{LQ}$,
which is conservative as it corresponds to the lower value of the theoretical cross section.
We find lower limits on M(LQ) at 235 GeV/c$^2$ ($\beta = 1$) and 176 GeV/c$^2$ ($\beta = 0.5$).
To obtain the best limit, we have combined the results from the two decay 
channels just described  with the result of a search for LQ in the
case where the LQ pair decays to neutrino and quark with branching ratio 
$ Br(LQ \to \nu q) = $1.0\cite{prl7200}. The individual channel analyses are in fact optimized for
fixed values of $\beta$ (1,0.5,0) while in the combined analysis, due to the contributions of the 
different decay channels, the signal acceptance can be naturally expressed as a function of $\beta$. 
As for the treatment of uncertainties, 
the searches in the $eejj$ and e$\nu jj$ channel use common criteria and sometimes apply the same kind of requirements 
so the uncertainties in the acceptances have been considered correlated. When calculating the limit 
combination including the $\nu\nu jj $ channel the uncertainties have been considered uncorrelated. 
For each $\beta$ value a 95\% C.L. upper 
limit on the expected number of events is returned for each mass, and by comparing this to the
theoretical expectation,  lower limits on the LQ mass are set.
The combined limit as a function of $\beta$ is shown in Figure 1, 
together with the individual channel limits. The combined mass limits are also tabulated in 
Table III.
\begin{table}
\caption{\label{tab:finalLimit}
95\% C.L. lower limits on the first generation scalar LQ mass (in GeV/c$^2$),
as a function of $\beta$. The limit from CDF\cite{d05} ($eejj$) Run I ($\sim 120 pb^{-1}$) is also given.}
\begin{ruledtabular}
\begin{tabular}{crcrcrc}
     $\beta$          & ee jj  & $e \nu jj$  &  $\nu \nu jj$   &   Combined  & CDF Run I \\
     \hline
     0.01             & $<$ 100      & $<$ 100          & 116             &   126       & -\\
     0.05             & $<$ 100      & $<$ 100          & 112             &   134       & - \\
     0.1              & $<$ 100      &  144             & $<$ 80          &   145       & - \\
     0.2              & $<$ 100      &  158             & $<$ 80          &   163       & - \\
     0.3              & 114          &  167             & $<$ 80          &   180       & - \\
     0.4              & 165          &  174             & $<$ 80          &   193       & - \\
     0.5              & 183          &  176             & $<$ 80          &   205       & - \\
     0.6              & 197          &  174             & $<$ 80          &   215       & - \\
     0.7              & 207          &  167             & $<$ 80          &   222       & - \\
     0.8              & 216          &  158             & $<$ 80          &   227       & - \\
     0.9              & 226          &  144             & $<$ 80          &   231       & - \\
     1.0              & 235          &  $<$100          & $<$ 80          &   236       & 213 \\
   \end{tabular}
\end{ruledtabular}
\end{table}
\begin{figure}
\includegraphics[scale=0.30]{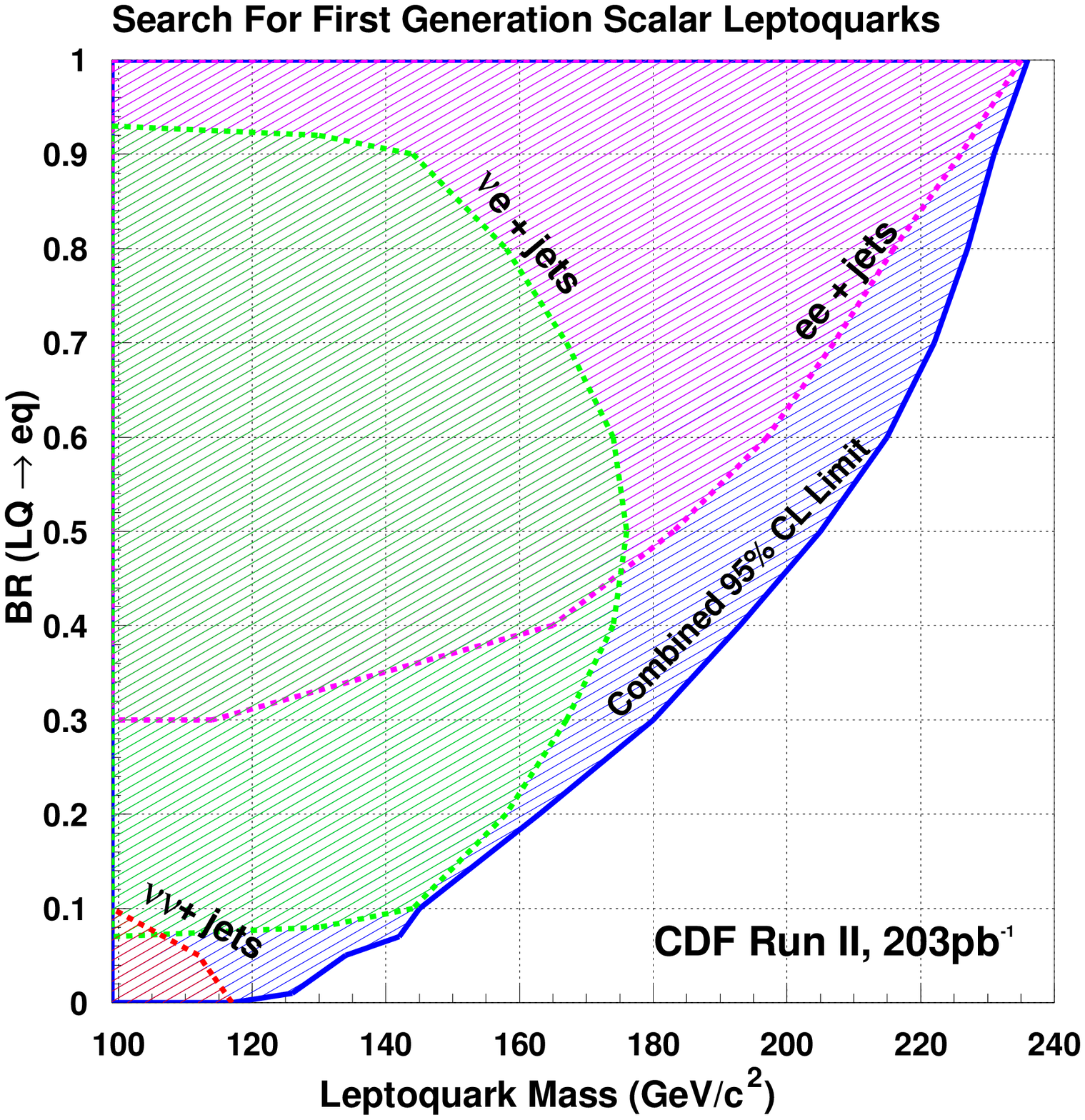} 
\caption{\label{fig:limit_combined}
LQ mass exclusion regions at 95\% C.L. as a function of Br(LQ$\to $~ eq).}
\end{figure}

In conclusion, we have performed a search for pair production of first generation
scalar LQs 
using 203 pb$^{-1}$ of proton-antiproton
collision data recorded by the CDF experiment during Run II of the Tevatron. 
The results from all the final state signatures are combined and no evidence of LQs production
is observed.
Assuming that a scalar LQ decays to electron and quark with variable branching ratio $\beta$
we exclude LQs with masses below 236 GeV/c$^2$ for $\beta$ = 1, 205 GeV/c$^2$ for $\beta$ = 0.5 
and 145 GeV/c$^2$ for $\beta$ = 0.1.

We thank the Fermilab staff and the technical staffs of the participating
institutions for their vital contributions. This work was supported by the
U.S. Department of Energy and National Science Foundation; the Italian
Istituto Nazionale di Fisica Nucleare; the Ministry of Education, Culture,
Sports, Science and Technology of Japan; the Natural Sciences and
Engineering Research Council of Canada; the National Science Council of the
Republic of China; the Swiss National Science Foundation; the A.P. Sloan
Foundation; the Bundesministerium fuer Bildung und Forschung, Germany; the
Korean Science and Engineering Foundation and the Korean Research
Foundation; the Particle Physics and Astronomy Research Council and the
Royal Society, UK; the Russian Foundation for Basic Research; the  Comisi\'on
Interministerial de Ciencia y Tecnolog\'{\i}a, Spain; and in part by the
European Community's Human Potential Programme under contract HPRN-CT-20002,
Probe for New Physics.


\begin{thebibliography}{99}
\bibitem{eta}
We use a cylindrical coordinate system around the beampipe in which $\theta$ is the polar angle,
$\phi$ is the azimuthal angle and $\eta = -ln (tan({\theta\over 2})) $. 
$E_T  = E \sin \theta $ and
$P_T =  P \sin \theta $, where $E$ is the energy measured by the calorimeter and $P$ the momentum measured
by the tracking system. $\bar { \MET}  = -\Sigma E^i_T \bar {n_i}$ where $\bar {n_i}$ 
is a unit vector that points from the
interaction vertex to the {\it i}th calorimeter tower in the transverse plane. $\MET$ is the magnitude
of $\bar{ \MET}$. $\MET$ is corrected following the correction of jet energies, and
if muons are identified in the event, $\MET$ is corrected for the muon momenta.
\end{thebibliography}
\end{document}